# Thermal design of focal plane assembly of wavefront camera for exoplanet imaging corona module

Chengcheng Wen[1,2,3], Shu Jiang[1,2], Mingming Xu[1,2,4], Lingyi Kong[1,2] and Jingning He[1,2]

[1]Nanjing Institute of Astronomical Optics & Technology, Chinese Academy of Sciences, Nanjing 210042, China

[2]CAS Key Laboratory of Astronomical Optics & Technology, Nanjing Institute of Astronomical Optics & Technology, Nanjing 210042, China

[3]University of Chinese Academy of Sciences, Beijing 100049, China

[4]mmxu@niaot.ac.cn

**Abstract.** The focal plane assembly of wavefront camera is the key imaging device of wavefront detection camera and the key load of the exoplanet imaging coronagraph module. In order to ensure the imaging quality and reduce the dark current and thermal noise, it is necessary to effectively dissipate the heat of the focal plane CCD and other high-power electronic devices to ensure the working performance and reliability of the focal plane assembly. Based on the limited space and cooling resources, this paper adopts the flexible graphene heat conducting cable and grooved heat pipes, and carries out detailed thermal design and thermal analysis of its cooling path. The finite element model is established by thermal analysis software and thermal simulation is carried out. in that steady state, the work temperature range of the CCD chip is -12.8~-10.9 °C, which meets the temperature control index, and the work temperatures of other components are also within the design requirements, which indicate that the thermal design scheme is reasonable and feasible, and meets the task requirements.

**Keywords:** The focal plane assembly, wavefront camera, thermal design, thermal analysis

## 1. Introduction

With the development of space technology, people began to explore the planets beyond the earth. Since the discovery of the "super-Earth" Kepler 62e, a number of terrestrial planets in the habitable zone have been discovered, bringing dawn to the search for extraterrestrial life and liveable planets. At the same time that the United States and Europe have successively launched space telescopes to explore exoplanets, China has also proposed the China Sky Survey Telescope (CSST) and the "Earth 2.0" program, which are equipped with space corona instruments, six 30cm aperture, 500 square degree wide-angle transiting telescopes and one 30cm aperture, 4 square degree micro-gravitational lens telescope respectively to complete the grand scientific task of exoplanet exploration [1].

Exoplanet imaging coronagraph module is one of the functional modules of China Survey Telescope (CSST), the first large space survey telescope in China. It undertakes the important mission of exploring exoplanets. is one of the functional modules of China Survey Telescope (CSST), the first large space survey telescope in China. It undertakes the important mission of exploring exoplanets. In order to improve that angular resolution of the coronagraph and reach the diffraction limit as far as possible, an adaptive optical system is introduced in the optical design to correct the wavefront distortion in real time. The adaptive optics system is mainly composed of wavefront detection, wavefront correction and wavefront control [2]. In a wavefront sensing camera, the wavefront distortion information collected by the micro-lens array is converted into electrical signals by the focal plane CCD component and transmitted to the wavefront controller, which directly determines the accuracy of wavefront sensing.

However, too high operating temperature will make CCD module produce dark current and thermal noise, reduce its photoelectric conversion ability, and seriously affect the quality of wavefront detection. Studies have shown that the dark current increases by an order of magnitude for every 7°C increase in the temperature level of the focal plane assembly [3]. Therefore, it is very important to control the working temperature of focal plane assembly by reasonable thermal design for exoplanet imaging detection.

At present, the thermal design of most space camera focal plane components abroad adopts the direct way of heat conduction aluminium strip or heat pipe and radiation heat dissipation plate to dissipate heat, or is directly exposed to the space environment, the outer surface is wrapped with multi-layer heat insulation components, and the refrigerator is used for active refrigeration under high temperature conditions, and the closed-loop active heating loop is used under low temperature conditions to maintain the focal plane temperature at a certain level. NASA's Earth Observation Satellite-1 (EO-1) carries an Advanced Land Imager (ALI), whose thermal control of the focal surface uses a thermally conductive aluminium strip to connect the focal surface and the heat dissipation surface, and a compensation heating device is set on the aluminium strip [4]. In China, for example, Tao Yang and Shilei Zhao of Beijing Institute of Space Mechatronics have designed a new type of loop heat pipe for simultaneous precision temperature control of multiple CCDs[5]. Liang Guo of Changchun Institute of Optics and Fine Mechanics of Chinese Academy of Sciences used phase change materials to reduce the working temperature level of CCD and improve its thermal stability [7][8]. In this paper, the flexible graphene heat conducting cable and heating pipes are used to establish an independent heat dissipation channel for CCD module, and the heat is conducted to the common heat dissipation plate to dissipate, so that the CCD temperature level meets the design index. Meanwhile, the stability of CCD working temperature is improved by the heat insulation measures such as multilayer heat insulation components and titanium alloy heat insulation pads. The use of flexible graphene also plays a role in damping. The rationality of our scheme is validated by the thermal analysis software.

## 2. Thermal control development task analysis

*2.1. Overview of wavefront sensing camera*

Structurally, the wavefront detection camera is composed of two parts, a focal plane box and a control box, which are connected by a conductive copper cable to form a flexible heat conduction connection, as shown in Figure 1. The focal plane box is installed on the optical substrate, and the control box is installed on the common heat dissipation plate. The overall installation diagram is shown in Figure 2.The focal plane box mainly carries a focal plane assembly for detecting wavefront tilt distortion and outputting a distorted image of an analog signal; The control box mainly carries electrical components such as signal processing components, bias and drive components, interconnection interface components and focal plane preamplifier circuit components, which are used to generate CCD drive signals and complete the preprocessing and transmission of distorted images. The body materials of the focal plane box and the control box are all made of aluminium alloy 2A12, which has the characteristics of low density, high strength, good processability and good thermal conductivity [6].

The focal plane module includes a frame transfer array EMCCD, image signal buffer circuit board and preamplifier circuit board. The structure is shown in Fig 3. The CCD is fixed with the focal plane box by a pressure plate, which is made of titanium alloy TC4 material, so that the CCD and the focal plane box are installed in a heat insulation manner; The flexible heat-conducting cable is installed with the box body, and the CCD is tightly attached to the flexible heat-conducting plate through the size precision control. the CCD image signal buff circuit board and that preamplifier circuit board are connected with the focal plane box body by bolt through a connecting plate.

The EMCCD has a one-stage thermoelectric cooler encapsulated on the bottom side of the interior with a maximum cooling temperature difference of 50°C. In order to ensure the cooling effect and reduce heat leakage, the CCD adopts Kovar alloy as the packaging material, and its thermal conductivity is as low as 15.5-17 $W/m·k$ [10]. In order to reduce its radiation heat leakage, its inner and outer surfaces are gold-plated.

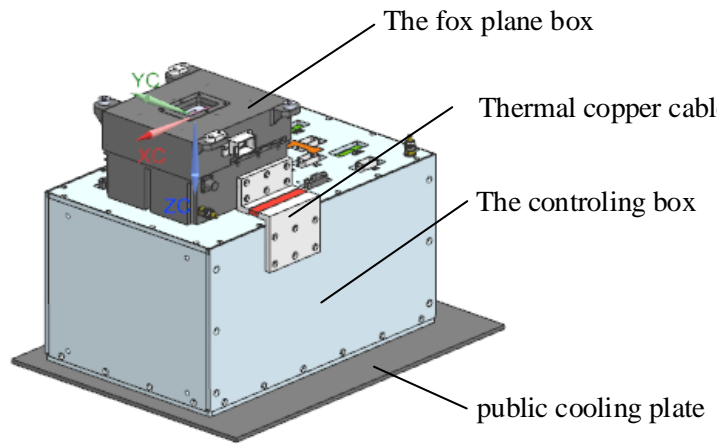

**Figure 1**. Structure diagram of wavefront detection camera

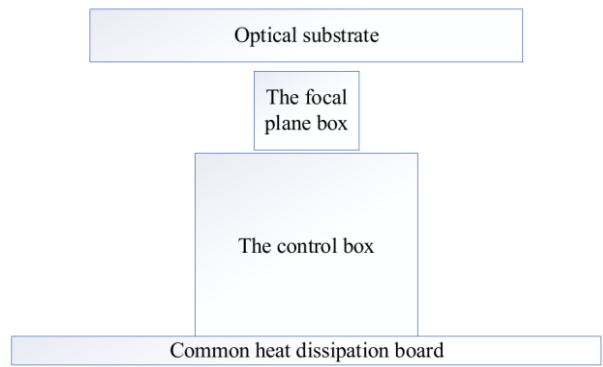

**Figure 2.** Schematic diagram of overall installation of wavefront detection camera.

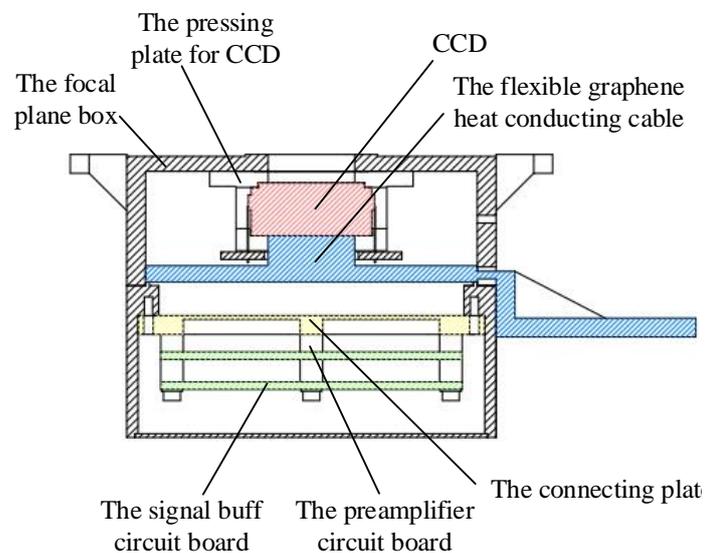

**Figure 3**. Structure diagram of focal plane assembly

*2.2. Thermal control task analysis*

In order to ensure the imaging quality of wavefront sensing camera and suppress dark current and thermal noise, the working temperature of CCD detector should be strictly controlled. The hot end of the thermoelectric cooler in the cooling process will produce a lot of waste heat. Therefore, the key to the whole thermal control task is: establish a reasonable heat dissipation channel to dissipate the waste heat of TEC hot end, so that the CCD can reach the best working temperature.

The EMCCD consumes 2.5W, and the internally integrated thermoelectric cooler consumes a maximum of 14.5W, for a total peak power consumption of 17W. The total power consumption of the image signal buffer board and the preamplifier board is 10W. The total power consumption of the components of the Wavefront Control Box is 28W. The working mode of the whole wavefront detection camera is long-time working, the temperature control index of the CCD is not higher than -5°C, the temperature control index of the focal plane box shell is below 55°C, and the junction temperature of other electronic devices is not higher than 80°C. The overall power consumption and temperature control indicators are shown in Table 1.

The whole module is placed in the cabin, and the common cooling plate is used as the whole cold source. The common cooling plate is connected with the evaporator of the loop heat pipe, and after the working medium of the loop heat pipe absorbs heat and evaporates in the evaporator, the heat is transmitted to the radiator, and finally the heat is discharged into the space through the radiator. Due to the influence of the heat flow outside the orbit, the non-uniformity of the heat consumption distribution of each single unit and the non-uniformity of the temperature caused by the heat conduction of the common heat dissipation plate, the temperature boundary at the heat pipe installation surface of the wavefront detection camera is +10°C under the high temperature condition and -20°C under the low temperature condition.

The small heat dissipation area and high heat flux density of TEC hot end bring some difficulties to thermal design. Due to the compact distance between the single unit, the temperature boundary of the wavefront sensing camera is also easily affected by the radiation of other single units under different working modes, which also brings great space restrictions to the thermal design.

**Table 1.** Summary table of work consumption, working mode and temperature control index of wavefront detection camera

| Component Name | Power (W) | Working mode | Temperature control index (°C) |
|---|---|---|---|
| CCD | 17 | Long-time working | ≤-5 |
| Preamplifier and signal buffer circuit board | 10 | Long-time working | ≤80 |
| Electronic components of the control box | 28 | Long-time working | ≤80 |

## 3. Discussion and analysis of overall scheme of thermal design

*3.1. Thermal design of heat conducting*

In order to prevent the influence of thermal coupling on CCD working temperature and improve the stability of it, the heat dissipation scheme of heat conduction copper plate and double heat pipe is adopted, and the heat dissipation channel is established separately to dissipate the heat from the TEC hot end to the common heat dissipation plate in time, so as to ensure the best cooling effect. The heat dissipation path is shown in Fig 4, and the heat is first diffused by the heat dissipation plate and then transferred to the common heat dissipation plate through the heat pipe. The contact surfaces are coated with GD414C thermally conductive silicone rubber to reduce thermal contact resistance. Graphene thermal cable has a high thermal conductivity on a plane, up to 4000-5000 $W/m·k$ [11]. After multi-layer film stacking, its equivalent thermal conductivity can reach 900-1500 $W/m·k$, and its flexibility also plays a role in vibration reduction. Aluminium-ammonia channel heat pipe is selected as heat pipe, and the heat transfer capacity of single heat pipe is 20W. Meanwhile, the shell of the focal plane box is connected with the optical substrate by a titanium alloy heat insulation pad, so that the thermal coupling between the optical substrate and the focal plane box is reduced.

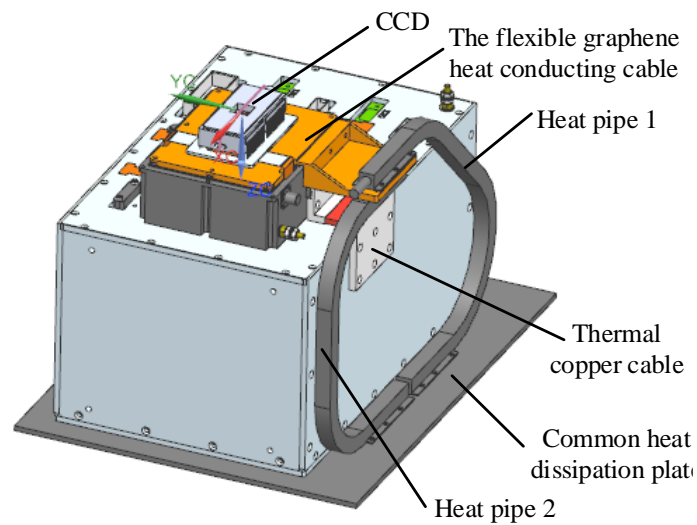

**Fig 4.** Schematic diagram of heat dissipation path of focal plane box

The image signal buffer circuit board and the preamplifier circuit board have a total heat consumption of 10W. The heat is conducted to the focal plane box through the connecting plate, and then transferred to the control box's shell through the heat-conducting copper cable, and finally dissipated through the common heat dissipation plate. The heat dissipation path is shown in Fig.4. The large power consumption devices on the circuit board are all designed with heat conducting copper strips connected with the frame for heat dissipation. The components are connected with the heat-conducting copper strip through the heat-conducting insulating pad and GD414C silicone rubber. The heat-conducting block is fixed on the frame with screws, and the contact surface is coated with heat-conducting grease.

*3.2. Thermal design of heat radiation*

The inner and outer surfaces of the coke box are blackened to enhance the radiation heat transfer with CCD. The outer surface of the focal plane box is wrapped with multi-layer thermal insulation components to isolate the influence of temperature fluctuations of other single units on the focal surface box. The multi-layer thermal insulation component consists of 10 units, the reflection layer is double-sided aluminized polyester film, the spacer layer is made of polyester mesh, and black polyimide film is selected as the inner and outer coating layer.

**4. Simulation analysis of thermal control scheme**

*4.1 Thermal resistance analysis of heat dissipation path of focal surface components*

The specific heat dissipation path of the focal surface component is as follows: the heat generated by the hot end of the CCD first passes through the flexible graphene heat conducting cable to the outside of the focal plane box, and then transfers to the common heat dissipation plate by two heat pipes; The heat generated by the pre-amplification and signal buffer circuit board is transferred from the connecting plate to the focal plane box wall, and then transferred from the copper heat conduction cable to the control box wall, and finally to the common heat dissipation plate. The thermal resistance analysis diagram of the overall heat dissipation path is shown in Fig 5. The thermal resistance between each thermal conductive component is the contact thermal resistance, and the rest is the thermal resistance of the thermal conductive component itself.

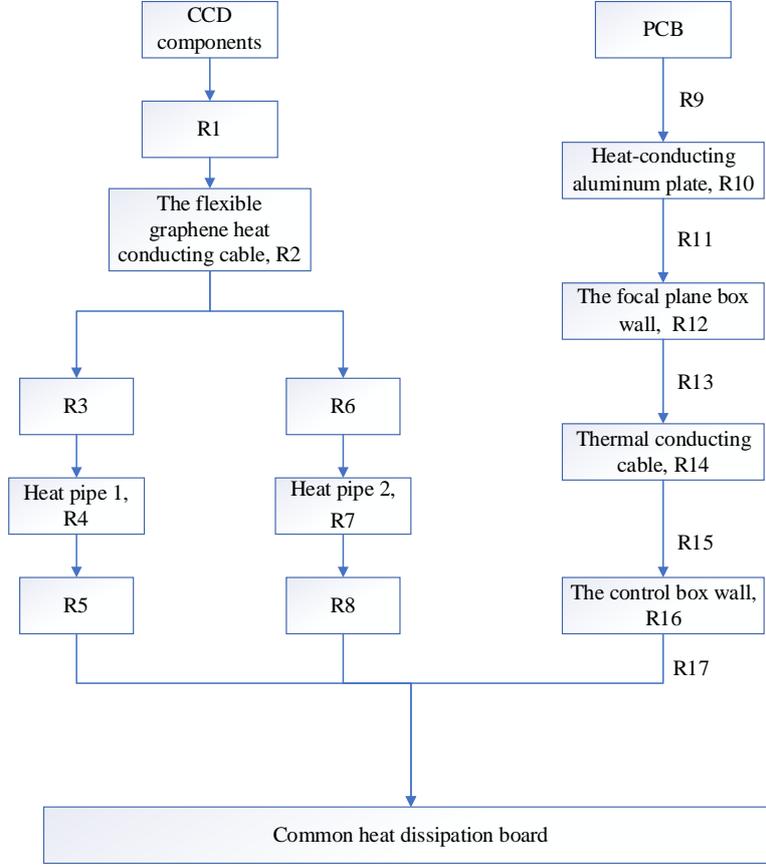

**Fig 5.** Analysis diagram of overall heat dissipation path's thermal resistance

According to the Fourier heat conduction law, the calculation formula of the thermal resistance of the heat conducting component is as follows, wherein $\Delta L$ is the characteristic length of the component in the heat transfer direction, $\lambda$ is the thermal conductivity of the component itself, and $A$ is the cross-sectional area in the heat transfer direction.

$$R_{self} = \Delta L(\lambda A)^{-1} \tag{1}$$

The thermal contact resistance between heat conducting components is calculated as follows, where $h$ is the contact heat transfer coefficient between surfaces.

$$R_{contact} = (hA)^{-1} \tag{2}$$

wherein the total thermal resistance of the heat transfer path of the CCD assembly is:

$$R_{CCD,total} = R_1 + R_2 + [(R_3 + R_4 + R_5)^{-1} + (R_6 + R_7 + R_8)^{-1}]^{-1} \tag{3}$$

The total thermal resistance of the remaining electronics heat transfer path is:

$$R_{rest,total} = R_9 + R_{10} + R_{11} + R_{12} + R_{13} + R_{14} + R_{15} + R_{16} + R_{17} \tag{4}$$

Finally, the theoretical temperature difference of the whole heat dissipation path can be calculated by using the following thermal resistance formula (5) and the corresponding boundary conditions, where in $\Delta T$ is the temperature difference and $Q$ is the heat flow.

$$\Delta T = QR \tag{5}$$

After the relevant parameters are taken into account, the total thermal resistance $R_{CCD,total} = 17.8℃/W$, and the total heat flow $Q = 17W$, is calculated. Consequently, the theoretical temperature difference $\Delta T_{theory}$ is calculated into formula (5):

$$\Delta T_{theory} = QR = 17 \times 1.78 = 30.26℃ \tag{6}$$

In vacuum, the radiative heat transfer between two contact surfaces is negligible. Heat transfer is mainly done by conduction of 0.1% of the actual contact area, which brings great thermal resistance [12]. As can be seen from Fig.4, the contact thermal resistance accounts for most of the overall thermal resistance. Therefore, GD414C silicone rubber or silicone rubber pad is applied on the contact surface in this design to reduce the contact thermal resistance, and the empirical value $h = 1500W/m^2 \cdot k$ is used for calculation.

*4.2 Thermal simulation analysis and result discussion*
*4.2.1. Finite element analysis.* According to the thermal design scheme described in Chapter 3, after simplification, the finite element model of focal plane assembly established by UG NX software is shown in Figure 6. The main steps of finite element analysis can be divided into: Model simplification, mesh generation, material property setting, thermal connection establishment, load setting, boundary condition and initial condition setting, and result analysis.

(1) *Model simplification and mesh generation*: The unimportant geometric features such as holes, fillets and stiffeners in the solid model are deleted, so the solid model is simplified. Therefore, the number of meshes is reduced, and the calculation difficulty of the subsequent solution process is decreased. In most of the models, 3D tetrahedral meshes are used. For thin shell structures such as common heat sink, 2D shell elements are used as simplification, and the element thickness is defined by equivalent volume method.

(2) *Material property setting*: Set the density, thermal conductivity, specific heat capacity, emissivity and other thermophysical parameters of materials used in the model. The thermophysical parameters of main structural materials used in the whole model are shown in Table 2. The multi-layer thermal insulation assembly is wrapped outside the focal plane box. The multi-layer thermal insulation assembly is equivalent to the thermal control coating in the simulation, and is thermally coupled with the external surface unit of the focal plane box in the form of 2D surface layer unit. According to the literature [9], the equivalent emissivity of multilayer insulation assembly is calculated as $\varepsilon_{equal} = 0.0382$. The equivalent thermal conductivity $\lambda_{equal} = 1000W/m^2 \cdot k$ is given to the flexible graphene heat conducting cable for thermal analysis calculation. The temperature difference between the two ends of the heat pipe is very small, so the equivalent thermal conductivity $\lambda_{equal} = 13000W/m^2 \cdot k$ is taken in the simulation.

(3) Establish thermal connection: Each heat conduction link is thermally coupled by contact thermal conductivity $h = 1500W/m^2 \cdot k$. The radiation heat transfer between surfaces is calculated by calculating the view angle coefficient of all surfaces in the heat radiation enclosure.

(4) Loads, boundary and initial conditions: The thermal loads of 17W and 28W were applied to the hot end surface of the focal plane box and the inner wall surface of the control box respectively. The radiant ambient temperature near the focal plane box and the initial temperature were both set to 20°C. Under high temperature condition, the common heat dissipation is 10°C as the thermal boundary.

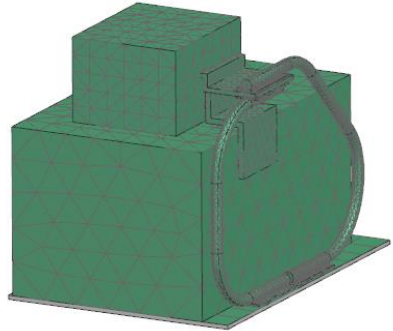

**Fig 6.** Finite element model

**Table 2.** Thermal analysis results of each component

| Materials | Thermal conductivity/ W.(m.K)$^{-1}$ | Emissivity | Density/ Kg.m$^{-3}$ | Specific heat capacity / J.(kg.K)$^{-1}$ |
|---|---|---|---|---|
| Aluminum alloy 2A12 (blacked) | 121 | 0.88 | 2800 | 921 |
| Titanium alloy TC4 | 6.8 | 0.1 | 4440 | 611 |
| Kovar alloy | 17 | 0.035 | 1800 | 1030 |
| Copper | 384 | 0.2 | 8920 | 385 |
| Graphene thermal cable | 1000 | 0.88 | 1800 | 1125 |

4.2.2. *Results and discussion*. The overall thermal analysis results of the wavefront camera are shown in Figure 7, and the results of the steady-state thermal analysis of the hot end of the CCD chip are shown in Figure 8. The maximum temperature is 39.1°C, the temperature of the contact surface with the boss of the CCD heat sink plate is about 35.6°C, and the average temperature is about 37.2°C. The working temperature range of CCD chip is -12.8~-10.9°C after being cooled by internal TEC thermoelectric cooler. The thermal analysis results of the temperature level of other parts of the focal plane assembly are shown in Table 3. It can be seen from the data in the table that the temperature level of each assembly meets the temperature control index.

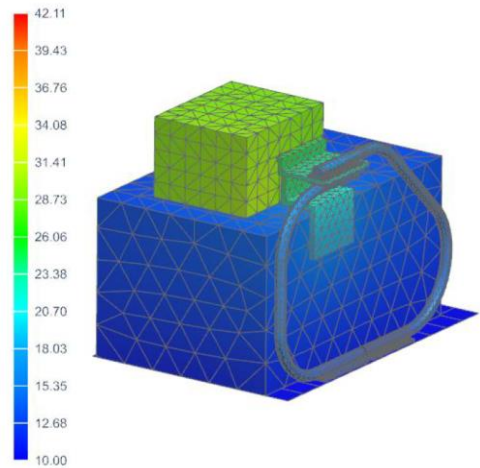

**Figure 7**. Overall temperature distribution nephogram of wavefront camera.

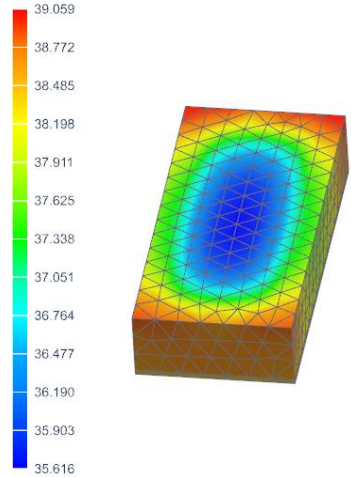

**Figure 8.** CCD package's temperature nephogram

**Table 3.** Thermal analysis results of each component

| Component name | Maximum temperature (°C) | Average temperature (°C) |
|---|---|---|
| CCD back package | 39.1 | 37.2 |
| Pre-amplification and signal buffer board | 41.9 | 39.1 |
| Pre-amplification circuit board | 40.6 | 37.2 |
| The shell of Focal plane box | 31.7 | 30.3 |

The simulation results show that the temperature difference from the CCD hot end to the common heat sink is $\Delta T_{simulation} = 27.2°C$   Tsimulation =27.2°C, which is in good agreement with the calculated temperature $\Delta T_{theory} = 30.26°C$. The error between the two results is only 11.25%, which preliminarily verifies the correctness of the thermal resistance analysis and calculation. The error sources may include: (1) the simplification error of the one-dimensional steady-state model; (2) The error caused by radiation heat transfer is not considered. As a consequence, the following methods to decrease the error are: (1) using a more accurate two-dimensional or three-dimensional steady-state heat conduction model; (2) The radiation heat transfer term is added to the heat balance differential equation.

The temperature difference between the hot and cold ends $\Delta T_{simulation} = 27.2°C$ is still large. In order to continuously reduce the working temperature of CCD, the results are analyzed and further optimized. According to the thermal resistance analysis in Section 4.1, the contact thermal resistance accounts for the vast majority of the total thermal resistance, which is the main reason for the large temperature difference. Therefore, improving the thermal contact resistance is the key to optimize this scheme. According to the contact formula (2), increasing the contact area and the contact heat transfer coefficient are the two ways to reduce the contact thermal resistance. The thermal contact resistance is related to the contact surface state, contact pressure, heat conduction filler and other factors. As a result, considering the design constraints of this scheme and the feasibility of optimization measures, the following improvement measures are obtained:

(1) Increase the size of that fin of the heat pipe to expand the contact area of both end of the heat pipe.

(2) Decrease the roughness of the contact surface and the pressure between the contact surface to enlarge the actual contact area.

(3) Replace Heat conductive silicone rubber with indium foil which has better thermal conductivity.

**5. Conclusions**

According to the characteristics and temperature control requirements of the focal plane box of the wavefront sensing camera, the thermal design and thermal analysis are carried out in detail. The flexible graphene heat conducting cable with high heat conductivity is adopted and connected with two parallel heat pipes, so that the heat dissipation is efficient and the vibration isolation is realized; An independent heat dissipation channel is designed for the CCD, and the stability of the working temperature of the CCD is ensured by the thermal protection effect of the titanium alloy and the multi-layer heat insulation component.

The results of theoretical calculation and finite element simulation analysis are in good agreement, and the error is 11.25%, which preliminarily verifies the feasibility of the scheme. In addition, through the analysis of thermal contact resistance, the further optimization measures are proposed. The thermal simulation results show that the working temperature range of CCD is -12.8 ~-10.9°C in steady state, which meets the temperature control index. The operating temperatures of other components are also within the design requirements. Therefore, the thermal design scheme is reasonable and feasible, which can effectively reduce the working temperature of CCD and provide a reference for the thermal control design of focal plane assembly.

**Reference**
[1] Lu P, Z, GAO T et al. Thermal control design and validation of high-resolution stereo mapping camera system. *Journal of Beijing University of Aeronautics and Astronautics*, 2023, 49(04):768-779.
[2] Yu Z, Meng QL Yu F et al. Design and validation of thermal control system for a low-orbit inclined orbit camera. Infrared and Laser Engineering, 2021, 50:149-154..
[3] Kong L, Jiang F, Wang JC et al. Thermal control of the CCD focal plane assembly of a microsatellite high-resolution camera. Spacecraft Recovery and Remote Sensing, 2023, 44:62-68.
[4] Wang DJ, Kong L, Zhang L, et al. Thermal design and verification of high-power focal plane of high-resolution camera. Spacecraft Environmental Engineering, 2023, 40:141-147.
[5] Yang T Zhao SL, Gao T, et al. Design and flight application of loop heat pipes for temperature control of dispersed heat sources in aerospace. Journal of Astronautics, 2021, 42:798-806.
[6] Guo N, Yu B, Xia CH, et al. High Temperature Stability Control of Focal Plane Stitching Base of Space Optical Camera. Spacecraft Recovery and Remote Sensing, 2020, 41:64-73.
[7] Guo L, Wu QW, Ding YL, et al. Design and verification of phase change temperature control of focal plane components of aerial cameras. Infrared and Laser Engineering, 2013, 42:2060-2067.
[8] Li YW, Yang HB, Zhang HW, et al. Application of phase change thermal control in CCD components of high-altitude optical remote sensors. Infrared and Laser Engineering, 2012, 41: 3016-3020.
[9] Wang D, Yan Y and Jin G. Thermal design and experimental research on high-speed TDICC focal plane components of space cameras. Opto-electronic Engineering, 2011, 38: 45-49.
[10] D. V. Dementev, T. Z. Lygdenova, and P. I. Kharlamov. Investigation and Optimization of a Cooling System Prototype for a Module of the Silicon Tracking System for the BM@N Experiment. Instruments and Experimental Techniques, 2021, 64: 40–47.
[11] Michel Engelhardt. Thermal Control of an Airborne Electronicsbay. American Institute of Aeronautics and Astronautics, 2007, p. 1217.
[12] Ren HY and Hu JG. Research progress on thermal contact resistance. Spacecraft Engineering, 1999, 8:47-57.